\documentclass[aps,pra,superscriptaddress,twocolumn,floatfix,showpacs]{revtex4-1}
\usepackage{graphicx,color}
\usepackage{amsmath,amssymb,bm}
\usepackage{hyperref}

\definecolor{darkred}{rgb}{0.90,0,0}
\definecolor{darkgreen}{rgb}{0,0.60,.2}
\definecolor{darkblue}{rgb}{0,0,1}
\definecolor{grey}{cmyk}{0,0,0,0.25}
\definecolor{orange}{cmyk}{0,0.6,0.8,0}

\begin{document}

\title{Particle statistics and lossy dynamics of ultracold atoms in optical lattices}

\author{J.~Yago Malo}
\affiliation{Department of Physics and SUPA, University of Strathclyde, Glasgow G4 0NG, Scottland, UK}

\author{E.~P.~L.~van Nieuwenburg}
\affiliation{Institute for Theoretical Physics, ETH Zurich, 8093 Zurich, Switzerland}
\affiliation{Institute for Quantum Information and Matter, Caltech, Pasadena, CA 91125, USA}

\author{M.~H.~Fischer}
\affiliation{Institute for Theoretical Physics, ETH Zurich, 8093 Zurich, Switzerland}

\author{A.~J.~Daley}
\affiliation{Department of Physics and SUPA, University of Strathclyde, Glasgow G4 0NG, Scottland, UK}

\date{\today}

\begin{abstract}
Experimental control over ultracold quantum gases has made it possible to investigate low-dimensional systems of both bosonic and fermionic atoms. In closed 1D systems there are a lot of similarities in the dynamics of local quantities for spinless fermions and strongly interacting ``hard-core'' bosons, which on a lattice can be formalised via a Jordan-Wigner transformation. In this study, we analyse the similarities and differences for spinless fermions and hard-core bosons on a lattice in the presence of particle loss. The removal of a single fermion causes differences in local quantities compared with the bosonic case, because of the different particle exchange symmetry in the two cases. We identify deterministic and probabilistic signatures of these dynamics in terms of local particle density, which could be measured in ongoing experiments with quantum gas microscopes. 
\end{abstract}


\maketitle

\section{Introduction}

In the past few years, there has been rapid progress in the characterization and control of dissipative dynamics for ultracold atoms in optical lattices. While these systems are most known for the possibility to engineer Hamiltonians for strongly interacting systems towards quantum simulation purposes, \cite{Bloch2012,Lewenstein2012}, the same level of microscopic understanding, in which models can be derived from first principles under well-controlled approximations, is also available for most of the dominant forms of dissipation that occur naturally in experiments. This applies, in particular, to our understanding of incoherent light scattering and the resulting dephasing of the many-body state \cite{Pichler2010, Sarkar2014}, and to our treatment of atom loss \cite{Barmettler2011}. Studying these sources of dissipation is of importance well beyond gaining a better understanding of experimental imperfections - it allows for the use of dissipation (i) in probing many-body states and their dynamics \cite{Muller20121,Vidanovic2014}, (ii) in the controlled preparation of interesting many-body states \cite{Muller20121,Kordas2012}, and (iii) in understanding how signatures of fundamental effects from closed systems (e.g., many-body localisation (MBL)) survive in the presence of coupling to an environment \cite{Luschen2017,Medvedyeva2016,Levi2016,Fischer2016,Nieuwenburg2017}. 

In this work, we explore how the differences between many-body states of hard-core bosons (HCB) and spinless fermions confined to move in one dimension (1D) can be probed using particle loss. In 1D, where strongly interacting bosons cannot pass each other, there are strong formal similarities between HCB and spinless fermions \cite{Girardeau1960}. These regimes have been realised in experiments with cold bosonic atoms in strongly confined 1D tubes \cite{Kinoshita2004}, and in lattices \cite{Stoferle2004,Paredes2004}, and the consequences can be seen clearly, even for just two atoms, in quantum gas microscope experiments \cite{Preiss2015}. For particles moving on a lattice, this similarity can be formalised via a Jordan-Wigner transformation to spin operators \cite{Sachdev2001}, where we see that for local models, the energy eigenvalues will be identical, and local correlations -- both for the eigenstates and out-of-equilibrium dynamics induced by changing local trap quantities -- will be equal as well. However, single-particle loss can generate differences in local quantities due to the different exchange symmetries in the many-body wavefunction. These differences manifest themselves in local density distributions, which are accessible with current experimental techniques in quantum gas microscopes\cite{Boll2016,Parsons2016,Cheuk2016,Brown2016,Haller2015,Edge2015}.

Making use of symmetries in tensor-network-based numerical methods, we calculate the dynamics of example systems for typical experimental sizes and parameter scales in the presence of loss. The efficient simulation of such systems requires the proper inclusion of symmetries in these numerical methods in order to account for the loss process in an affordable manner. We first study the loss process as a deterministic event and then employ a quantum trajectory approach \cite{Dum1992,Molmer1993,Daley2014} to determine features of bosons and fermions that survive stochastically occurring loss events. 

The rest of the paper is structured as follows. In Sec.~II, we discuss the theoretical model for fermions and bosons confined in 1D subject to dissipation. In Sec.~III, we highlight the differences we expect to observe between the different types of particle statistics in the event of a loss, and in Sec.~IV, we describe the numerical approach that allows for the computationally efficient simulation of a system subject to this kind of dissipation. In Sec.~V, we then analyze the dynamics following losses that occur at deterministic times and locations, identifying accessible parameter regimes where the differences between HCB and spinless fermions are significant and could be engineered and observed using quantum gas microscopes  \cite{McKay2011,Weitenberg2011,Gericke2008}. In Sec.~VI, we study which of those features identified in Sec.~V survive under stochastic losses and which of them vanish when the losses occur randomly, providing local and spatially averaged quantities that can be obtained through density measurements. Finally, in Sec.~VII we discuss our findings.

\section{Model: Fermions and hard-core bosons in the presence of local particle loss}

In this section, we introduce a model for particle loss in spinless fermions or hard-core bosons confined to move along one direction of an optical lattice (and tightly confined in the other two directions). 

For fermions in the lowest Bloch band of the optical lattice, the system is well described by a tight-binding Hamiltonian($\hbar\equiv 1$),
\begin{equation}
\hat{H}=-J\sum_{\left\langle ij\right\rangle}\,\hat{a}_{i}^{\dagger}\hat{a}_{j}\,,
\label{eq:ham1}
\end{equation}
where $\left\langle ij\right\rangle$ indicates that the sum runs over all nearest neighbours, the operator $\hat{a}^{(\dagger)}_{i}$ annihilates (creates) a fermionic particle on the site $i$ where $\hat{n}_{a,i}=\hat{a}_{i}^{\dagger}\hat{a}_{i}\in[0,1]$ is the fermionic number operator of the site $i\in[1,M]$, where $M$ is the lattice system size, and $J$ is the tunneling amplitude in the lattice. The fermionic operators obey the usual anticommutation rules, $\left\{ \hat{a}_{i}^{(\dagger)},\,\hat{a}_{j}^{(\dagger)}\right\} =0;\,\left\{ \hat{a}_{i},\,\hat{a}_{j}^{\dagger}\right\} =\delta_{i,j}$.

An analogous model can be considered for the case of hard-core bosons, for which the Hamiltonian can be seen as a limiting case of the Bose-Hubbard Hamiltonian \cite{Gersch1963}, and is given by:
\begin{equation}
\hat{H}=-J\sum_{\left\langle ij\right\rangle}\,\hat{b}_{i}^{\dagger}\hat{b}_{j},\,\,\, \hat{b}_l^2\equiv 0\,,
\label{eq:ham2}
\end{equation}
where the operator $\hat{b}^{(\dagger)}_{i}$ annihilates (creates) a bosonic particle on the site $i$ and $\hat{n}_{b,i}=\hat{b}_{i}^{\dagger}\hat{b}_{i}\in[0,1]$ is the bosonic number operator for the site $i$. In contrast to the fermionic case, the bosonic creation/annihilation operators obey usual commutation rules, $\left [ \hat{b}_{i}^{(\dagger)},\,\hat{b}_{j}^{(\dagger)}\right ] =0;\,\left [ \hat{b}_{i},\,\hat{b}_{j}^{\dagger}\right ] =\delta_{i,j}$.

 \begin{figure}[bt]
  \centering%
  \includegraphics[width=1\columnwidth]{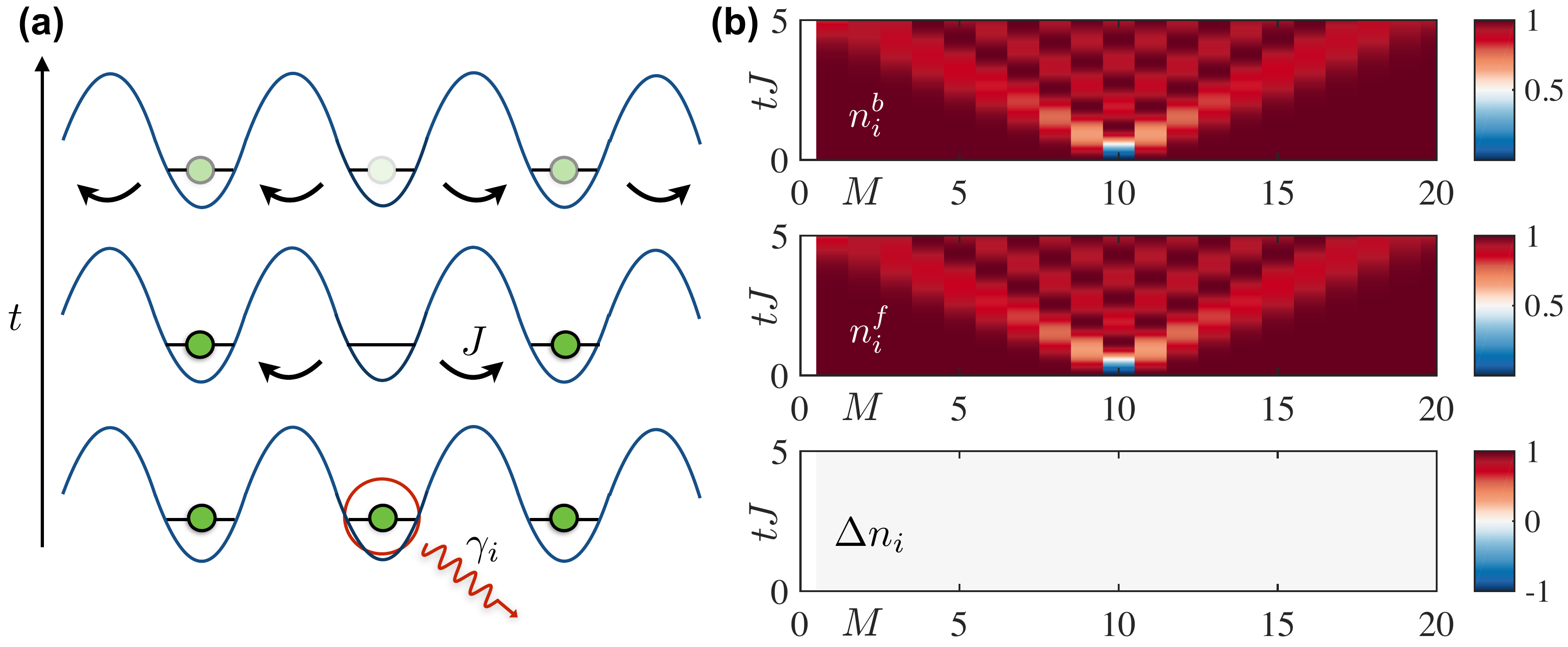} 
  \caption{(a) Diagram of a loss event in an optical lattice on site $i$ with probability $\gamma_i$. The density hole created will propagate through normal tunneling processes and will delocalize over time; (b) Evolution of the particle density for bosons $\hat{n}^b_i$, fermions $\hat{n}^f_i$ and the normalised difference of these, $\Delta \hat{n}_i$, as a function of time. In this case, loss occurs on site $i=10$ on a lattice with $M=20$ from an initial product state with a single particle on each site, so a single sign is applied to the fermionic wavefunction and both profiles remain identical, i.e. $\Delta \hat{n}_i=0$.}
  \label{fig:Fig_0}
 \end{figure}

We can describe the dissipative dynamics of such systems in the presence of particle loss via a master equation for the system density operator $\rho_{tot}$. The master equation arises on a microscopic level because in these atomic physics systems we can usually make a Born-Markov approximation, justified by the existence of a single dominant frequency for each process (given by the energy of the lost atom for single-particle loss, and by the photon frequency for dephasing due to light scattering \cite{Daley2014}). The resulting master equation is given by
\begin{equation}
\frac{d\rho}{dt}=-i\left[\hat{H},\,\rho\right]-\frac{1}{2}\sum_{m}^{2M}\gamma_{m}(\hat{J}_{m}^{\dagger}\hat{J}_{m}\rho+\rho \hat{J}_{m}^{\dagger}\hat{J}_{m}-2\hat{J}_{m}\rho \hat{J}_{m}^{\dagger})\,,
\label{eq:master}
\end{equation}
where $\hat{J}_m=\hat{a}_{m} (\hat{b}_{m})\, {\rm{for}} \,m\in[1,M]$ represents the loss of a fermion (boson) on site $m$, $\hat{J}_m=\hat{n}_{a,m} (\hat{n}_{b,m}) \,{\rm{for}} \, m\in[M+1,2M]$ describes the dephasing process and $\gamma_m$ is the decay amplitude for the $m$-th dissipation channel that will be different for dephasing and loss processes. The inclusion of dephasing, which is naturally present in experimental realisations due to light scattering \cite{Pichler2010, Poletti2013, Sarkar2014, Luschen2017}, will allow us to test whether any differences between spinless fermions and bosons are diminished by this form of dissipation. Numerical solutions to the evolution of the system will be discussed in subsequent sections. 

Fig.~\ref{fig:Fig_0} shows a schematic view of the density profiles after a deterministic loss event in the middle site at $t=0$, beginning from an initial product state with one atom on every lattice site. Because of the simple initial state and the single loss process, the density distributions for bosons and fermions as a function of time are identical, i.e., the normalised difference,
\begin{equation}
\Delta \hat{n}_i=\frac{\hat{n}_i^{b}-\hat{n}_i^{f}}{\hat{n}_i^{b}+\hat{n}_i^{f}},
\end{equation}
where $\hat{n}_i^{b}=\langle \hat{n}_{b,i}\rangle$ and $\hat{n}_i^{f}=\langle \hat{n}_{a,i}\rangle$, is zero in this case. In analysing different parameter regimes and identifying differences between HCB and spinless fermions, we focus particularly on this quantity in the following sections.

 \section{Differences between fermions and bosons in the presence of loss}
 
Bosonic and fermionic atoms will behave differently in the presence of dissipation as a result of the difference in the sign of the wavefunction under exchange of particles. One way to see this is to use a Jordan-Wigner transformation to map each of these cases to spin operators \cite{Sachdev2001}. A single-species model for hard-core bosons can be directly rewritten as an equivalent spin-1/2 model, with the spin states associated with each lattice site denoting presence ($|\!\!\uparrow\rangle$) or absence ($|\!\!\downarrow\rangle$) of a particle on that site. Because bosons commute, the mapping between particle annihilation operators and spin lowering operators $\hat{\sigma}^-_l$ is a direct replacement, $\hat{b}_l \rightarrow \hat{\sigma}_l^-$.  However, the same mapping for fermions requires a sign determined by a string operator in order to account for anti-commutation of the annihilation operators with all other operators present in the state description, $\hat{a}_l\rightarrow (-1)^{\sum_{i<l}\hat{n}_{a,i}}\hat{\sigma}^-_l$. 

It is clear that a loss event can thus affect the many-body state differently for fermions and for hard-core bosons. Our goal here is to identify whether there are differences that can be extracted solely from the local density distribution, $\hat{n}_{a/b,l}$, which translates the same way into spin operators for bosons and fermions under a Jordan-Wigner transformation, $\hat{n}_{a/b,l} \rightarrow \hat{\sigma}_l^+ \hat{\sigma}_l^-$. Indeed, for unitary dynamics involving only onsite and nearest-neighbour terms, the two cases, of spinless fermions and HCB, are identical as all of the signs vanish.

The vanishing of these phases for fermions after the transformation can be easily understood if we consider that they arise in the first place due to the commutation of the annihilation operators with the rest of the operators describing the state of the system. The local density is proportional to a product of two operators $\hat{n}_{a,i}=\hat{a}^{\dagger}_l\hat{a}_l $, thus any phase that arises from the commutation will cancel and $\hat{n}_{a,i}=\hat{\sigma}_l^+ \hat{\sigma}_l^-$. Similarly, if we consider terms that only include first-neighbour tunneling $\hat{a}^{\dagger}_l\hat{a}_{l\pm1}$ all signs will disappear; and so under local perturbations, the dynamics are identical for both species. Physically, this arises because these local operators cannot (for spinless fermions or HCB) exchange two particles that are present on different sites. However, in the presence of loss, there is an additional sign from the commutation of the operator to the respective site. This also implies that (at least in this formalism) the loss operator is in principle non-local for fermions.

\section{Numerical methods and the relevance of system symmetries}

In order to determine the dynamics for up to tens of lattice sites (which correspond to current experiments \cite{Boll2016,Parsons2016,Cheuk2016,Brown2016,Haller2015,Edge2015}), we make use of tensor-network methods \cite{White1992,Verstraete2008, Schollwock2011}. These methods provide us with efficient tools to compute the time evolution of both closed and open 1D many-body systems through the time-evolving block decimation (TEBD) algorithm \cite{Vidal2003}. In particular, open dynamics have been described through tensor networks by mapping the density operator $\rho$ to a matrix product operator (MPO) \cite{Pirvu2010,Verstraete2004}. Alternatively, the system evolution can be computed using a quantum trajectory approach \cite{Daley2014} where we can map the density operator dynamics to a stochastic sampling of pure-state evolutions in the form of matrix product states (MPS).

When we consider the case of fermionic losses, the string operator $\hat{N}_{<k}=(-1)^{\sum_{i<k}\hat{n}_{a,i}}$ is an expensive operator to compute in terms of matrix product states, as it is a highly non-local term and lacks a simple representation as an MPO. However, as shown recently \cite{Nieuwenburg2017}, this operator can be efficiently applied if we split our state representation into parity conserving sectors. In a similar manner, we will benefit from making use of number conserving sectors \cite{Schollwock2005,Daley2004}, which optimise time evolution calculations for pure states implementing quantum-trajectories techniques for the master equation \cite{Daley2014}. 

In this particular case, we structure the matrix product state in such a way that the storage scheme for the local tensor $A^{d_i}$ (with maximum bond dimension $D$) for site $i$ with local dimension $d_i$ (in our case $d_i=\mathrm{dim}(n_i)=2$) groups together the states that correspond to every possible population quantum number to the left of site $i$. In this way, the string operator reduces to a trivial value $N_{<k}=\pm 1$ depending on a number that we store for every state in every site. As a result, the application of an annihilation operator, representing a loss in the lattice, becomes the application of a local operator multiplied by a known phase.

Note that all the other terms appearing in the dynamics [Eqs.(\ref{eq:ham1})-(\ref{eq:master})], both in the unitary and the dissipative part, are either proportional to $\hat{n}_{a,i}=\hat{a}^{\dagger}_i\hat{a}_i$,  or proportional to $\hat{a}^{\dagger}_i\hat{a}_{i\pm1}$, with all string operators evaluating to one as discussed in Sec.~III. Thus, the only non-local phase arises from the loss term that we have already adapted. As a result, we can apply standard TEBD algorithms to compute the time evolution and study the dissipative dynamics through quantum trajectories efficiently as all our terms become local.

Below we will first use these techniques to compute the dynamics resulting from loss at a particular site and a particular time. We then follow this by simulating a master equation that describes loss processes that occur at random during the dynamics.

\section{Deterministic losses}

In this section we study the dynamics of the system when we induce the loss of a particle starting from a particular initial state. This could be achieved in a quantum gas microscope using single site addressing (freezing the state by rapidly increasing the lattice depth, changing the internal state, and removing the resulting atoms \cite{Weitenberg2011}), or by making use of addressing with an electron beam \cite{Gericke2008}. 

Outside of the loss events, we compute the unitary evolution of spinless fermions and hardcore bosons governed by the Hamiltonians in Eq.~(\ref{eq:ham1}) and Eq.~(\ref{eq:ham2}). We first consider the atoms to be in a product state and induce a loss at $t=0$ on site $M_0=M/2$. A second loss event is then induced at a chosen time $t=\tau_0$ on site $M_0-\delta_M$, with $\delta_M$ a chosen lattice distance. We consider different filling factors $n_0=N_0/M$, where $N_0$ is the initial number of particles and $M$ is the number of lattice sites. In particular, we will start both with a configuration consisting of a single atom per site ($n_0=1$) and a charge density wave state, with only odd sites occupied initially ($n_0=0.5$).

 \begin{figure}[tb]
  \centering%
  \includegraphics[width=1\columnwidth]{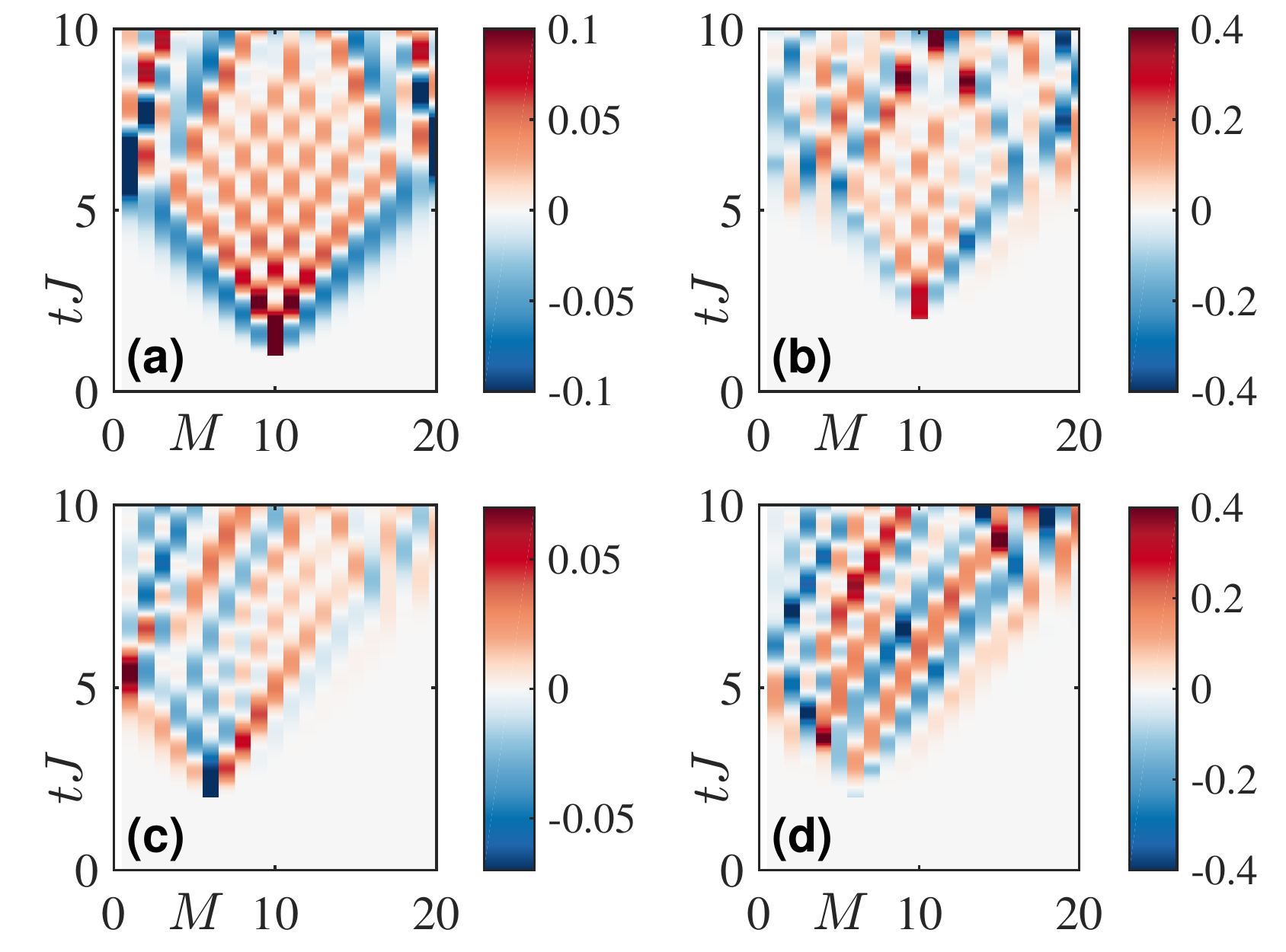} 
  \caption{(a) Evolution of the difference in density distribution $\Delta \hat{n}_i$ for a system with $M=20$, $n_0=1$, $D=100$, $dt=0.001$, $J=1$, $\epsilon_i=0\,(\forall i)$, $\tau_0=1$, $\delta_M=0$; (b) Same as (a) with $\tau_0=2$ and $n_0=0.5$; (c) Same as (a) with $\tau_0=2$ and $\delta_M=4$; (d) Same as (a) with $\tau_0=2$, $n_0=0.5$ and $\delta_M=4$. These calculations are performed beginning from a product state with the corresponding densities indicated above: a single particle on each lattice site ($n_0=1$) or every odd site ($n_0=0.5$). }
  \label{fig:Fig_1}
 \end{figure}
 
In Fig.~\ref{fig:Fig_1}, we present the difference in density distribution $\Delta \hat{n}_i$. We observe that the dynamical evolution of the density of hardcore bosons and fermions is identical up to the point of the second loss event. This occurs because the initial product state results in a single phase being applied to the whole fermionic state $N_{<M/2}=\pm 1$, as was shown in Fig.~1. However, when the second loss occurs, the delocalisation of the initial hole results in a superposition of different numbers of particles to the left of any given site, and so the effect of the phase is non-trivial. As a result, the densities of HCBs and spinless fermions start to differ in a well-defined light-cone in a ballistic manner. This is reminiscent of the spreading of correlation functions we expect in this system \cite{Lieb1972}. In the unit-filling regime (Fig.~\ref{fig:Fig_1}a and Fig.~\ref{fig:Fig_1}c), we observe that only losses occurring close to the region where the first one occurred ($\delta_M\sim 0$), i.e. where the population is not still deeply in the unit-filling Mott phase, lead to a significant difference between bosons and fermions, as it is only in this case that the effects of the string operator are non-trivial. In the case of half filling (Fig.~\ref{fig:Fig_1}b and Fig.~\ref{fig:Fig_1}d), as the particles are allowed to quickly delocalize, the difference is greater in magnitude and the relevance of the position, where the second loss occur, disappears.

 \begin{figure}[bt]
  \centering%
  \includegraphics[width=1\columnwidth]{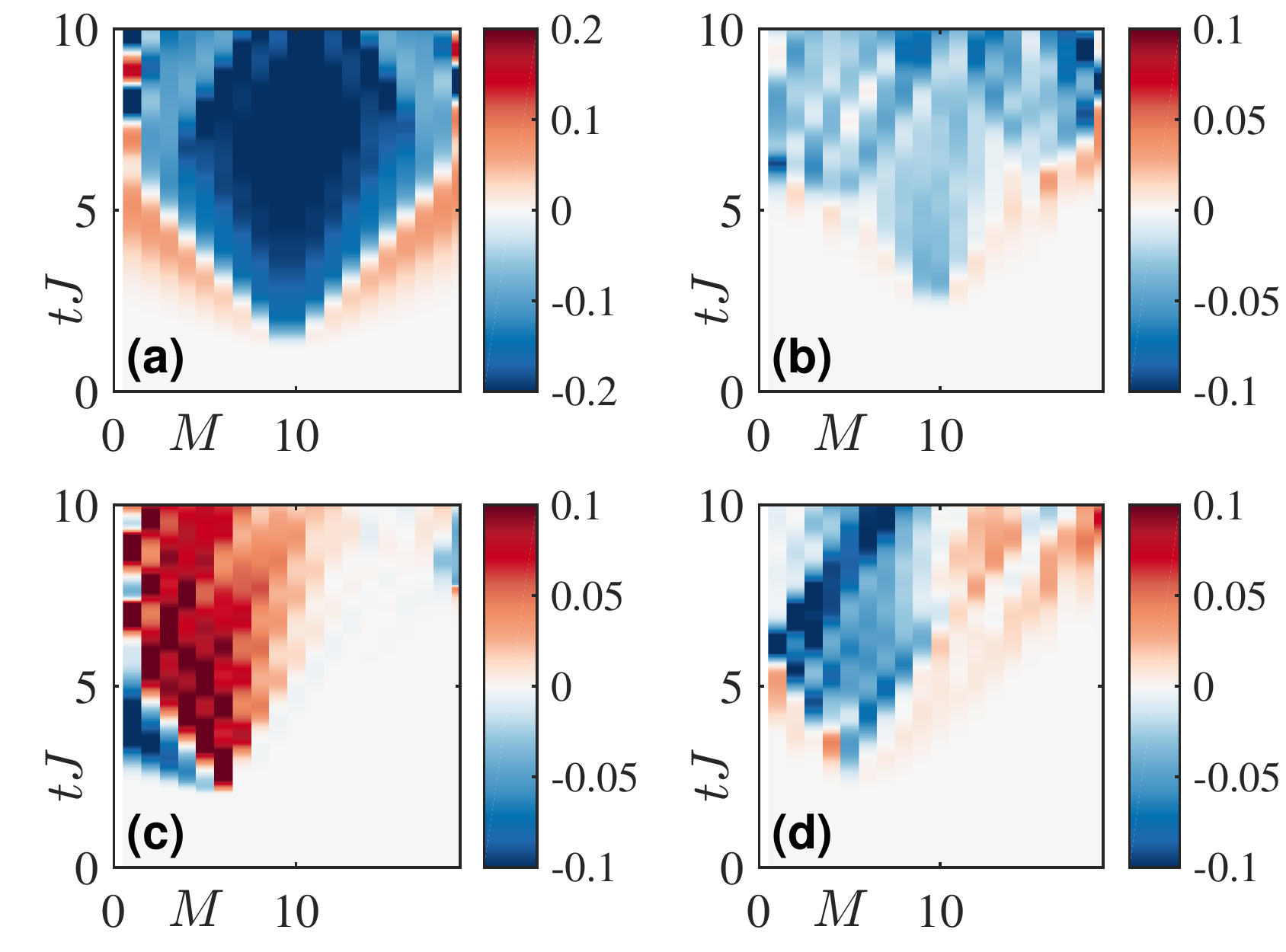} 
  \caption{(a) Evolution of the weighted difference in the entanglement entropy $\Delta S$ at every lattice bipartition for a system with $M=20$, $n_0=1$, $D=100$, $dt=0.001$, $J=1$, $\epsilon_i=0\,(\forall i)$, $\tau_0=1$, $\delta_M=0$; (b) Same as (a) with $\tau_0=2$ and $n_0=0.5$; (c) Same as (a) with $\tau_0=2$ and $\delta_M=4$; (d) Same as (a) with $\tau_0=2$, $n_0=0.5$ and $\delta_M=4$. These calculations are performed beginning from a product state with the corresponding densities indicated above: a single particle on each lattice site ($n_0=1$) or every odd site ($n_0=0.5$).}
  \label{fig:Fig_2}
 \end{figure}
 
In Fig.~\ref{fig:Fig_2}, we present the weighted difference in the entanglement entropy $\Delta S=\frac{S_{vN}^{b}-S_{vN}^{f}}{S_{vN}^{b}+S_{vN}^{f}}$, where $S_{vN}^{b/f}=-\textrm{tr}(\rho^{b/f}\ln\rho^{b/f})$ at every bipartition of both the bosonic and fermionic systems. This is another indicator of the differences in the dynamics, and can also be measured directly in quantum gas microscope experiments for both fermions and HCBs \cite{Islam2015,Daley2012,Pichler2013b}. After the losses occur we observe regions with higher entropy for the bosonic case as the non-local phase associated with the fermionic loss permits a faster spreading of the entanglement along the system. Note that now lattice configurations away from unit filling (compare Fig.~\ref{fig:Fig_2}a and Fig.~\ref{fig:Fig_2}b) exhibit smaller differences between fermions and bosons. This is due to the fact that the higher mobility in the lattice contributes to overall higher values of $S_{vN}$ for both species and we are representing normalized differences. Similar to the case of the density, losses that occur near the boundary of the lattice (Fig.~\ref{fig:Fig_2}c,d) lead to a smaller observable difference as the fermionic state is closer to a product state.

\section{Non-deterministic losses}

While these differences between HCBs and spinless fermions can be probed directly in experiments by inducing losses at particular lattice sites and times, it is important also to ask whether the difference is directly observable when losses occur at random, for example, via collisions with background gas or photon scattering bursts \cite{Luschen2017}. In experiments, we also typically deal with two other elements that we have not included up to now. At finite interaction strengths between bosons, a nearest-neighbour interaction term arises in second-order perturbation theory, which we model by considering offsite interactions of the form  $\sum_{\langle ij\rangle}U\hat{n}_i\hat{n}_j$. Also, we usually encounter some level of dephasing due to light scattering.

To properly investigate the effects of the latter in typical experiments, we compute the dissipative dynamics in the presence of both losses (with amplitude $\gamma_l$) and dephasing (with amplitude $\gamma_d$) for the same initial configurations provided in the deterministic case. We focus our interest again on quantities related to local densities that can be measured in quantum gas microscopes, and which would be identical for HCBs and spinless fermions in the absence of losses. 

 \begin{figure}[tb]
  \centering
  \includegraphics[width=1\columnwidth]{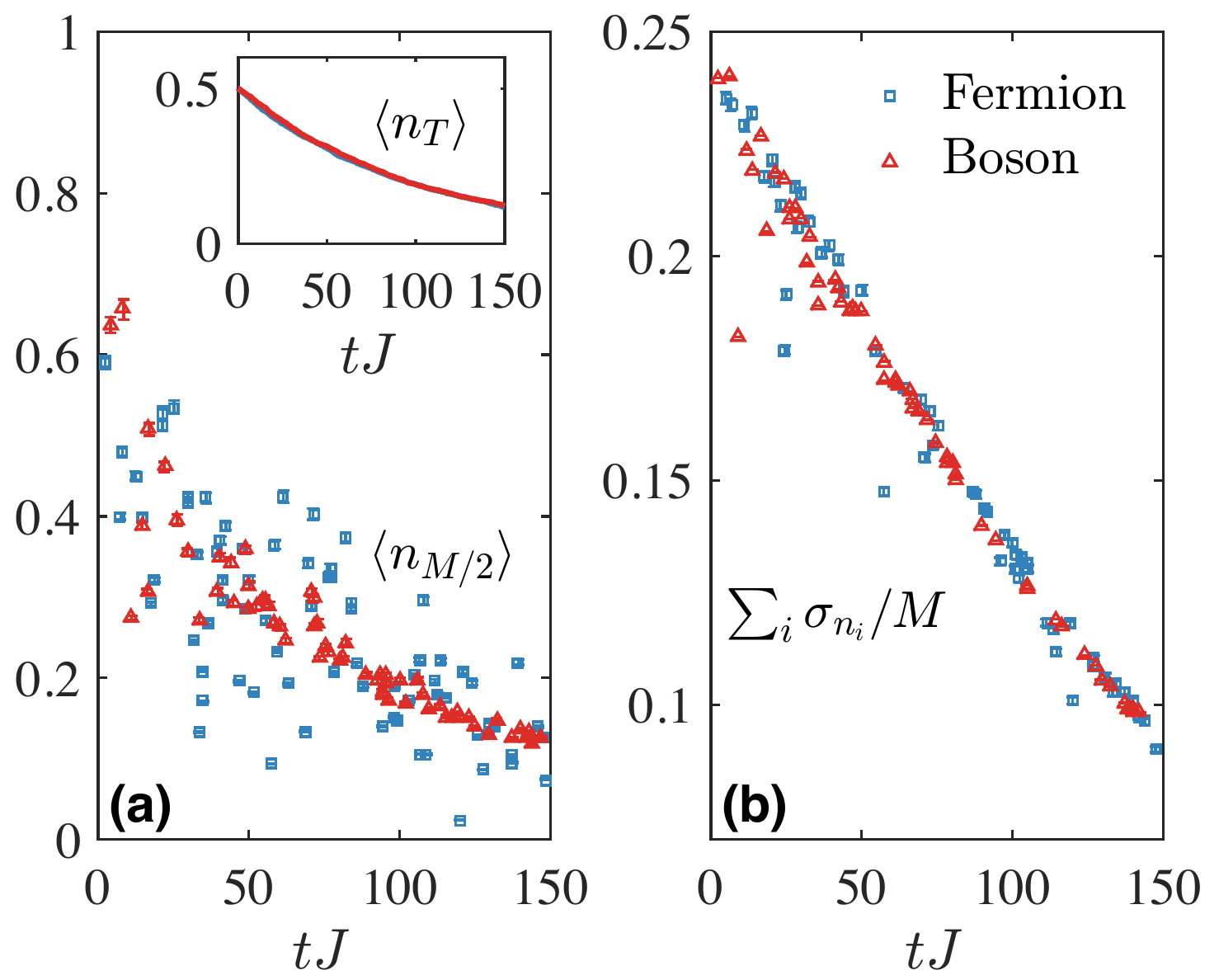} 
  \caption{(a) Comparison of bosonic and fermionic evolution of the middle site density $\langle n_{M/2} \rangle$ for a system with $M=16$, $n_0=0.5$, $J=1$, $\gamma_l=0.01$, $\gamma_d=0$; numerical parameters are $dt=0.001$ and $D=200$. The inset shows the total particle number $n_T$ to provide some guidance over the evolution of the total occupation in the lattice as losses occur. (b) Same as (a) for the normalized total density fluctuations $\sum_i \sigma_{n_i}/M$. These calculations are performed beginning from a charge density wave at half-filling, with a particle on each even-numbered site. Note that these functions are rapidly oscillating, and that each point represents a snapshot of the values on a randomly spaced grid in time. The data includes statistical error bars, which are contained within the point markers in most of the cases.}
  \label{fig:Fig_3}
 \end{figure}

In Fig.~\ref{fig:Fig_3}, we plot the evolution of different observables for both fermions and bosons. Fig.~\ref{fig:Fig_3}a shows the local density on the central site $\langle n_{M/2} \rangle$, where the values for the two species are measurably different. Specifically, we observe that local densities experience significantly larger amplitude oscillations in the fermionic case, and a substantial difference in the density profile persists over time. In Fig.~\ref{fig:Fig_3}b, we compute the lattice average fluctuations $\sum_i\sigma_{n_i}=\sum_i(\langle \hat{n}_i^2 \rangle - \langle \hat{n}_i \rangle ^2)$. The first conclusion that we can extract is that while fluctuations in local values appear to differ quite strongly, lattice averaging seems to wash out the discrepancy. The simplest example of an averaged quantity is the mean particle number per site, which is shown in the inset of Fig.~\ref{fig:Fig_3}a, and simply corresponds to an exponential decay with a rate $\gamma_l$. Although the total fluctuations presented in Fig.\ref{fig:Fig_3}b are very similar, a particularly precise measurement could distinguish these, as the fermionic case has a higher oscillation amplitude. However, to provide more realistic conditions for measurements we consider a set of snapshots at randomized time steps that mimic the measurement of local densities in a quantum gas microscope. We observe that at certain times, the measurement coincides with a peak in the fermionic oscillations leading to a clear distinction but overall the differences remain much smaller for this lattice-averaged density fluctuation. 
 \begin{figure}[tb]
  \centering
  \includegraphics[width=1\columnwidth]{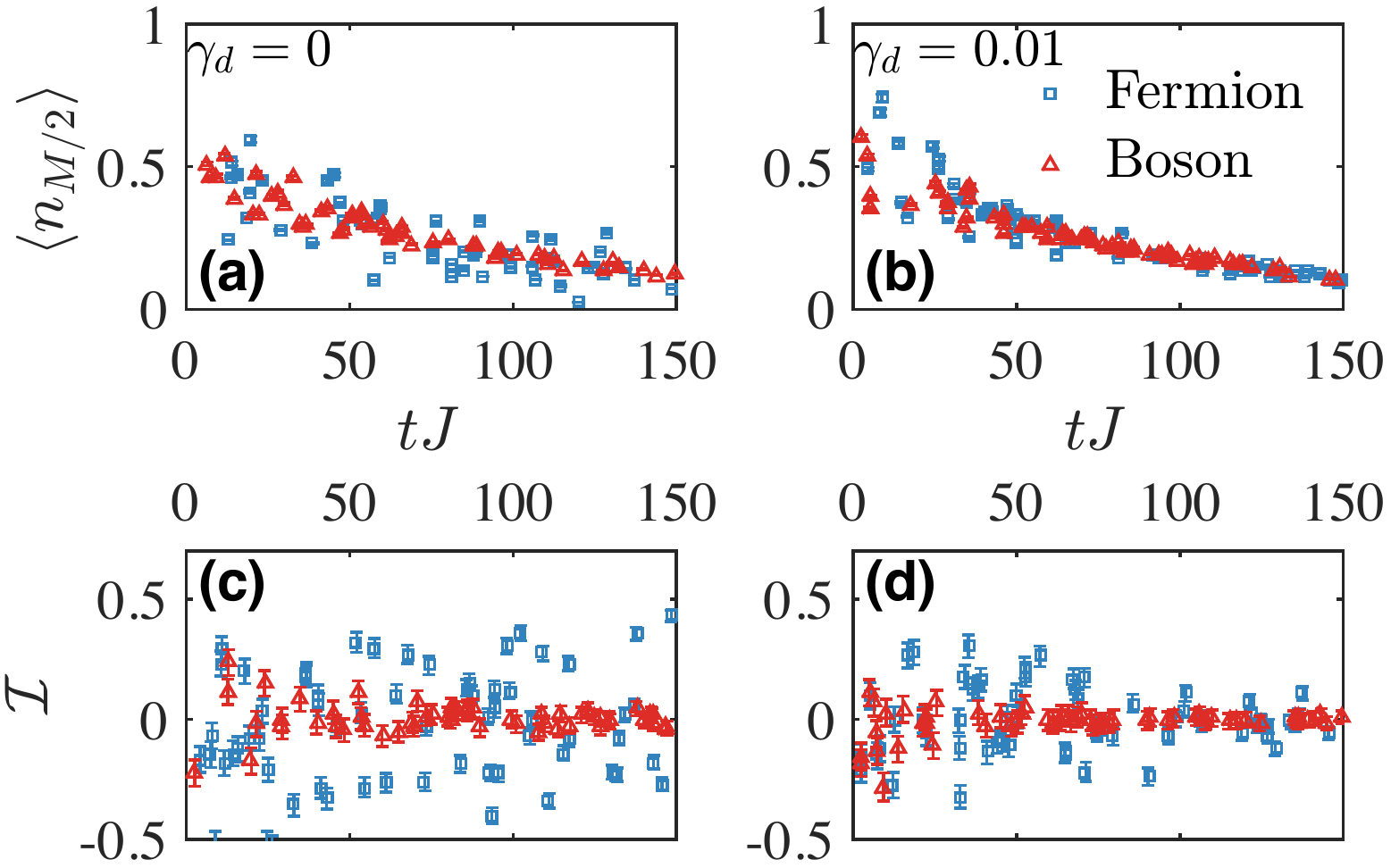} 
  \caption{(a) Comparison of bosonic and fermionic evolution of the middle site density $\langle n_{M/2} \rangle$ for a system with $M=16$, $n_0=0.5$, $J=1$, $dt=0.001$, $\gamma_l=0.01$, $\gamma_d=0$; (b) same as (a) with $\gamma_d=0.01$; (c) Comparison of bosonic and fermionic evolution of the imbalance $\mathcal{I}$ with same parameters as (a); (d) same as (c) with $\gamma_d=0.01$. These calculations are performed beginning from a charge density wave at half-filling, with a particle on each even-numbered site. Note that these functions are rapidly oscillating, and that each point represents a snapshot of the values on a regularly spaced grid in time.}
  \label{fig:Fig_4}
 \end{figure}
 
However, not all the global quantities suffer from the averaging. To look at this further, we consider the total odd-even site density imbalance, 
\begin{equation}
\mathcal{I}=\frac{\hat{n}^{o}-\hat{n}^{e}}{\hat{n}^{o}+\hat{n}^{e}},
\end{equation}
where $\hat{n}^{o/e}=\sum^M_{i\in odd/even} \langle \hat{n}_i \rangle$; the imbalance is a commonly considered variable in the context of many-body localisation in cold atoms \cite{Schreiber2015}.
In Fig.~\ref{fig:Fig_4}, we show both the local density on the central lattice site and the system imbalance. We also analyze the robustness of both quantities in the presence of dephasing. In the absence of this source of dissipation, both quantities allow us to differentiate between bosonic and fermionic dynamics as the profiles are significantly separated. However, while the differences in the local densities reduce at longer times, the fermionic imbalance exhibits much larger oscillations than the bosonic one and this feature persists over the simulated length of time. We observe that the inclusion of dephasing, corresponding to the values shown in Fig.~\ref{fig:Fig_4}b,d, reduces this clear separation.  Nevertheless, this reduction is much stronger in the local density, whereas the even-odd imbalance seems to be more robust to the presence of dephasing.
 
 \begin{figure}[tb]
  \centering
  \includegraphics[width=1\columnwidth]{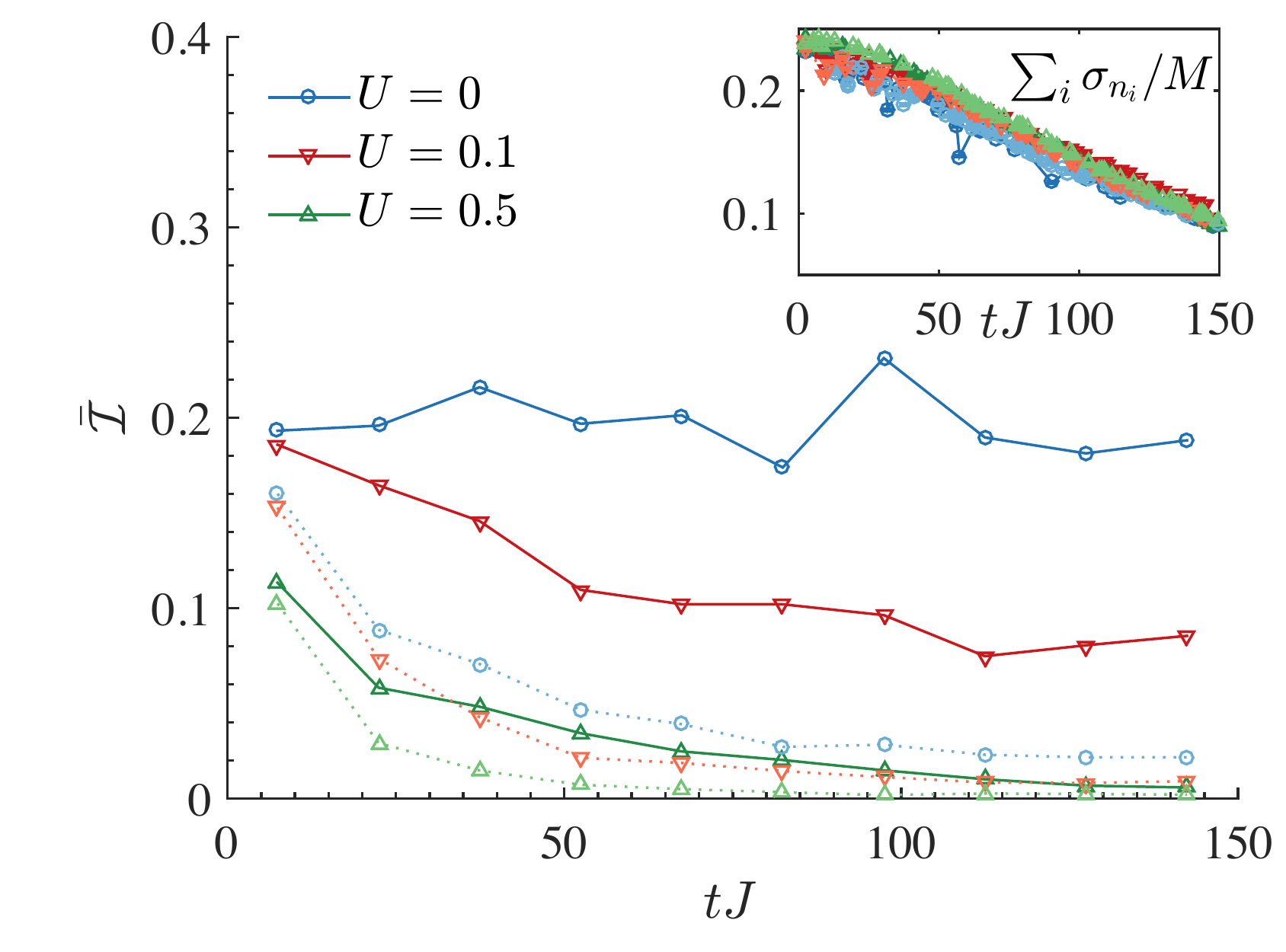} 
  \caption{Comparison of bosonic (dashed line) and fermionic (solid line) evolution of the time-block averaged imbalance $\bar{\mathcal{I}}=\sum_{i}^{i+N_{\Delta t}}|\mathcal{I}(t_i)|/N_{\Delta t}$ for a system with $M=16$, $n_0=0.5$, $J=1$, $dt=0.001$, $N_{\Delta t}=300$, $\gamma_l=0.01$, $\gamma_d=0$ and variable offsite interaction strength $U$. The imbalance average drops over time in the presence of interaction but remain distinguishable for both species. Inset: total density fluctuation for the same parameters, included for the purpose of comparison. Here, all lines overlap while we observe relevant differences in the imbalance. These calculations are performed beginning from a charge density wave at half-filling, with a particle on each odd-numbered site.}
  \label{fig:Fig_6}
 \end{figure}
 
Finally, in Fig.~\ref{fig:Fig_6} we investigate whether this imbalance discrepancy remains robust in the presence of offsite interactions. As the imbalance is a highly-oscillating function specially for the fermionic case, we present here a time-block averaged imbalance $\bar{\mathcal{I}}=\sum_{i}^{i+N_{\Delta t}}|\mathcal{I}(t_i)|/N_{\Delta t}$, where $N_{\Delta t}$ is the number of time points over which we average. For the sake of clarity, the absolute value is required as the imbalance should average to zero after a transient time much shorter than the timescale we simulate. As the interaction ramps up, the separation reduces between the fermionic and bosonic case. Nevertheless, the separation is still much greater than the one we can observe from the total fluctuations of the density. From this analysis we can establish that the imbalance -- a global quantity related to local densities -- is robust to moderate interactions and to moderate dephasing at rates comparable to the losses, and provides an interesting quantity with which to investigate differences between HCBs and spinless fermions also in the case of randomized losses in space and time.

 \section{Summary and outlook}

In this article, we have investigated how quantities that are related to the local density, and are experimentally measurable in quantum gas microscopes, allow us to distinguish between spinless fermions and hard-core bosons in the presence of particle loss. In the absence of loss, these quantities would in each case be identical for fermions and bosons. We have shown that the understanding of loss is not only a relevant element towards the correct description of the experimental conditions, but it can also play an essential role as a tool to access information about aspects of the closed-system dynamics. 

In the future, understanding these processes could help probe particular types of many-body effects. It is an important ingredient to better understand the effects of losses in optical lattice experiments, as well as to investigate the effects of losses in the study of systems with slow intrinsic time scales, e.g., many-body localised states in the presence of dissipation.

\begin{acknowledgements}
We thank Anton Buyskikh for helpful discussions. Work at the University of Strathclyde was supported by the EPSRC Programme Grant
DesOEQ (EP/P009565/1), by the European Union Horizon 2020 collaborative
project QuProCS - Quantum Probes for Complex Systems (grant agreement
641277), and by the EOARD via AFOSR grant number FA2386-14-1-5003. Results were obtained using the EPSRC funded ARCHIE-WeSt High Performance Computer (www.archie-west.ac.uk). EPSRC grant no. EP/K000586/1.
 
\end{acknowledgements}

\bibliographystyle{apsrev}
\bibliography{fermionic_loss_bib}

\begin{thebibliography}{46}
\expandafter\ifx\csname natexlab\endcsname\relax\def\natexlab#1{#1}\fi
\expandafter\ifx\csname bibnamefont\endcsname\relax
  \def\bibnamefont#1{#1}\fi
\expandafter\ifx\csname bibfnamefont\endcsname\relax
  \def\bibfnamefont#1{#1}\fi
\expandafter\ifx\csname citenamefont\endcsname\relax
  \def\citenamefont#1{#1}\fi
\expandafter\ifx\csname url\endcsname\relax
  \def\url#1{\texttt{#1}}\fi
\expandafter\ifx\csname urlprefix\endcsname\relax\def\urlprefix{URL }\fi
\providecommand{\bibinfo}[2]{#2}
\providecommand{\eprint}[2][]{\url{#2}}

\bibitem[{\citenamefont{Bloch et~al.}(2012)\citenamefont{Bloch, Dalibard, and
  Nascimbene}}]{Bloch2012}
\bibinfo{author}{\bibfnamefont{I.}~\bibnamefont{Bloch}},
  \bibinfo{author}{\bibfnamefont{J.}~\bibnamefont{Dalibard}}, \bibnamefont{and}
  \bibinfo{author}{\bibfnamefont{S.}~\bibnamefont{Nascimbene}},
  \bibinfo{journal}{Nat Phys} \textbf{\bibinfo{volume}{8}},
  \bibinfo{pages}{267} (\bibinfo{year}{2012}).

\bibitem[{\citenamefont{Lewenstein et~al.}(2012)\citenamefont{Lewenstein,
  Sanpera, and Ahufinger}}]{Lewenstein2012}
\bibinfo{author}{\bibfnamefont{M.}~\bibnamefont{Lewenstein}},
  \bibinfo{author}{\bibfnamefont{A.}~\bibnamefont{Sanpera}}, \bibnamefont{and}
  \bibinfo{author}{\bibfnamefont{V.}~\bibnamefont{Ahufinger}},
  \emph{\bibinfo{title}{Ultracold Atoms in Optical Lattices: Simulating quantum
  many-body systems}} (\bibinfo{publisher}{OUP Oxford}, \bibinfo{year}{2012}).

\bibitem[{\citenamefont{Pichler et~al.}(2010)\citenamefont{Pichler, Daley, and
  Zoller}}]{Pichler2010}
\bibinfo{author}{\bibfnamefont{H.}~\bibnamefont{Pichler}},
  \bibinfo{author}{\bibfnamefont{A.~J.} \bibnamefont{Daley}}, \bibnamefont{and}
  \bibinfo{author}{\bibfnamefont{P.}~\bibnamefont{Zoller}},
  \bibinfo{journal}{Phys. Rev. A} \textbf{\bibinfo{volume}{82}},
  \bibinfo{pages}{063605} (\bibinfo{year}{2010}).

\bibitem[{\citenamefont{Sarkar et~al.}(2014)\citenamefont{Sarkar, Langer,
  Schachenmayer, and Daley}}]{Sarkar2014}
\bibinfo{author}{\bibfnamefont{S.}~\bibnamefont{Sarkar}},
  \bibinfo{author}{\bibfnamefont{S.}~\bibnamefont{Langer}},
  \bibinfo{author}{\bibfnamefont{J.}~\bibnamefont{Schachenmayer}},
  \bibnamefont{and} \bibinfo{author}{\bibfnamefont{A.~J.} \bibnamefont{Daley}},
  \bibinfo{journal}{Phys. Rev. A} \textbf{\bibinfo{volume}{90}},
  \bibinfo{pages}{023618} (\bibinfo{year}{2014}).

\bibitem[{\citenamefont{Barmettler and Kollath}(2011)}]{Barmettler2011}
\bibinfo{author}{\bibfnamefont{P.}~\bibnamefont{Barmettler}} \bibnamefont{and}
  \bibinfo{author}{\bibfnamefont{C.}~\bibnamefont{Kollath}},
  \bibinfo{journal}{Phys. Rev. A} \textbf{\bibinfo{volume}{84}},
  \bibinfo{pages}{041606} (\bibinfo{year}{2011}).

\bibitem[{\citenamefont{M{\"u}ller et~al.}(2012)\citenamefont{M{\"u}ller,
  Diehl, Pupillo, and Zoller}}]{Muller20121}
\bibinfo{author}{\bibfnamefont{M.}~\bibnamefont{M{\"u}ller}},
  \bibinfo{author}{\bibfnamefont{S.}~\bibnamefont{Diehl}},
  \bibinfo{author}{\bibfnamefont{G.}~\bibnamefont{Pupillo}}, \bibnamefont{and}
  \bibinfo{author}{\bibfnamefont{P.}~\bibnamefont{Zoller}},
  \bibinfo{journal}{Advances In Atomic, Molecular, and Optical Physics}
  \textbf{\bibinfo{volume}{61}}, \bibinfo{pages}{1 } (\bibinfo{year}{2012}),
  ISSN \bibinfo{issn}{1049-250X}, \bibinfo{note}{advances in Atomic, Molecular,
  and Optical Physics}.

\bibitem[{\citenamefont{Vidanovi\ifmmode~\acute{c}\else \'{c}\fi{}
  et~al.}(2014)\citenamefont{Vidanovi\ifmmode~\acute{c}\else \'{c}\fi{}, Cocks,
  and Hofstetter}}]{Vidanovic2014}
\bibinfo{author}{\bibfnamefont{I.}~\bibnamefont{Vidanovi\ifmmode~\acute{c}\else
  \'{c}\fi{}}}, \bibinfo{author}{\bibfnamefont{D.}~\bibnamefont{Cocks}},
  \bibnamefont{and}
  \bibinfo{author}{\bibfnamefont{W.}~\bibnamefont{Hofstetter}},
  \bibinfo{journal}{Phys. Rev. A} \textbf{\bibinfo{volume}{89}},
  \bibinfo{pages}{053614} (\bibinfo{year}{2014}).

\bibitem[{\citenamefont{Kordas et~al.}(2012)\citenamefont{Kordas, Wimberger,
  and Witthaut}}]{Kordas2012}
\bibinfo{author}{\bibfnamefont{G.}~\bibnamefont{Kordas}},
  \bibinfo{author}{\bibfnamefont{S.}~\bibnamefont{Wimberger}},
  \bibnamefont{and} \bibinfo{author}{\bibfnamefont{D.}~\bibnamefont{Witthaut}},
  \bibinfo{journal}{EPL (Europhysics Letters)} \textbf{\bibinfo{volume}{100}},
  \bibinfo{pages}{30007} (\bibinfo{year}{2012}).

\bibitem[{\citenamefont{L\"uschen et~al.}(2017)\citenamefont{L\"uschen, Bordia,
  Hodgman, Schreiber, Sarkar, Daley, Fischer, Altman, Bloch, and
  Schneider}}]{Luschen2017}
\bibinfo{author}{\bibfnamefont{H.~P.} \bibnamefont{L\"uschen}},
  \bibinfo{author}{\bibfnamefont{P.}~\bibnamefont{Bordia}},
  \bibinfo{author}{\bibfnamefont{S.~S.} \bibnamefont{Hodgman}},
  \bibinfo{author}{\bibfnamefont{M.}~\bibnamefont{Schreiber}},
  \bibinfo{author}{\bibfnamefont{S.}~\bibnamefont{Sarkar}},
  \bibinfo{author}{\bibfnamefont{A.~J.} \bibnamefont{Daley}},
  \bibinfo{author}{\bibfnamefont{M.~H.} \bibnamefont{Fischer}},
  \bibinfo{author}{\bibfnamefont{E.}~\bibnamefont{Altman}},
  \bibinfo{author}{\bibfnamefont{I.}~\bibnamefont{Bloch}}, \bibnamefont{and}
  \bibinfo{author}{\bibfnamefont{U.}~\bibnamefont{Schneider}},
  \bibinfo{journal}{Phys. Rev. X} \textbf{\bibinfo{volume}{7}},
  \bibinfo{pages}{011034} (\bibinfo{year}{2017}).

\bibitem[{\citenamefont{Medvedyeva et~al.}(2016)\citenamefont{Medvedyeva,
  Prosen, and \ifmmode \check{Z}\else \v{Z}\fi{}nidari\ifmmode~\check{c}\else
  \v{c}\fi{}}}]{Medvedyeva2016}
\bibinfo{author}{\bibfnamefont{M.~V.} \bibnamefont{Medvedyeva}},
  \bibinfo{author}{\bibfnamefont{T.~c.~v.} \bibnamefont{Prosen}},
  \bibnamefont{and} \bibinfo{author}{\bibfnamefont{M.}~\bibnamefont{\ifmmode
  \check{Z}\else \v{Z}\fi{}nidari\ifmmode~\check{c}\else \v{c}\fi{}}},
  \bibinfo{journal}{Phys. Rev. B} \textbf{\bibinfo{volume}{93}},
  \bibinfo{pages}{094205} (\bibinfo{year}{2016}).

\bibitem[{\citenamefont{Levi et~al.}(2016)\citenamefont{Levi, Heyl, Lesanovsky,
  and Garrahan}}]{Levi2016}
\bibinfo{author}{\bibfnamefont{E.}~\bibnamefont{Levi}},
  \bibinfo{author}{\bibfnamefont{M.}~\bibnamefont{Heyl}},
  \bibinfo{author}{\bibfnamefont{I.}~\bibnamefont{Lesanovsky}},
  \bibnamefont{and} \bibinfo{author}{\bibfnamefont{J.~P.}
  \bibnamefont{Garrahan}}, \bibinfo{journal}{Phys. Rev. Lett.}
  \textbf{\bibinfo{volume}{116}}, \bibinfo{pages}{237203}
  (\bibinfo{year}{2016}).

\bibitem[{\citenamefont{Fischer et~al.}(2016)\citenamefont{Fischer, Maksymenko,
  and Altman}}]{Fischer2016}
\bibinfo{author}{\bibfnamefont{M.~H.} \bibnamefont{Fischer}},
  \bibinfo{author}{\bibfnamefont{M.}~\bibnamefont{Maksymenko}},
  \bibnamefont{and} \bibinfo{author}{\bibfnamefont{E.}~\bibnamefont{Altman}},
  \bibinfo{journal}{Phys. Rev. Lett.} \textbf{\bibinfo{volume}{116}},
  \bibinfo{pages}{160401} (\bibinfo{year}{2016}).

\bibitem[{\citenamefont{van Nieuwenburg et~al.}(2017)\citenamefont{van
  Nieuwenburg, Malo, Daley, and Fischer}}]{Nieuwenburg2017}
\bibinfo{author}{\bibfnamefont{E.~P.} \bibnamefont{van Nieuwenburg}},
  \bibinfo{author}{\bibfnamefont{J.~Y.} \bibnamefont{Malo}},
  \bibinfo{author}{\bibfnamefont{A.~J.} \bibnamefont{Daley}}, \bibnamefont{and}
  \bibinfo{author}{\bibfnamefont{M.~H.} \bibnamefont{Fischer}}
  (\bibinfo{year}{2017}).

\bibitem[{\citenamefont{Girardeau}(1960)}]{Girardeau1960}
\bibinfo{author}{\bibfnamefont{M.}~\bibnamefont{Girardeau}},
  \bibinfo{journal}{Journal of Mathematical Physics}
  \textbf{\bibinfo{volume}{1}}, \bibinfo{pages}{516} (\bibinfo{year}{1960}).

\bibitem[{\citenamefont{Kinoshita et~al.}(2004)\citenamefont{Kinoshita, Wenger,
  and Weiss}}]{Kinoshita2004}
\bibinfo{author}{\bibfnamefont{T.}~\bibnamefont{Kinoshita}},
  \bibinfo{author}{\bibfnamefont{T.}~\bibnamefont{Wenger}}, \bibnamefont{and}
  \bibinfo{author}{\bibfnamefont{D.~S.} \bibnamefont{Weiss}},
  \bibinfo{journal}{Science} \textbf{\bibinfo{volume}{305}},
  \bibinfo{pages}{1125} (\bibinfo{year}{2004}), ISSN \bibinfo{issn}{0036-8075}.

\bibitem[{\citenamefont{St\"oferle et~al.}(2004)\citenamefont{St\"oferle,
  Moritz, Schori, K\"ohl, and Esslinger}}]{Stoferle2004}
\bibinfo{author}{\bibfnamefont{T.}~\bibnamefont{St\"oferle}},
  \bibinfo{author}{\bibfnamefont{H.}~\bibnamefont{Moritz}},
  \bibinfo{author}{\bibfnamefont{C.}~\bibnamefont{Schori}},
  \bibinfo{author}{\bibfnamefont{M.}~\bibnamefont{K\"ohl}}, \bibnamefont{and}
  \bibinfo{author}{\bibfnamefont{T.}~\bibnamefont{Esslinger}},
  \bibinfo{journal}{Phys. Rev. Lett.} \textbf{\bibinfo{volume}{92}},
  \bibinfo{pages}{130403} (\bibinfo{year}{2004}).

\bibitem[{\citenamefont{Paredes et~al.}(2004)\citenamefont{Paredes, Widera,
  Murg, Mandel, Folling, Cirac, Shlyapnikov, Hansch, and Bloch}}]{Paredes2004}
\bibinfo{author}{\bibfnamefont{B.}~\bibnamefont{Paredes}},
  \bibinfo{author}{\bibfnamefont{A.}~\bibnamefont{Widera}},
  \bibinfo{author}{\bibfnamefont{V.}~\bibnamefont{Murg}},
  \bibinfo{author}{\bibfnamefont{O.}~\bibnamefont{Mandel}},
  \bibinfo{author}{\bibfnamefont{S.}~\bibnamefont{Folling}},
  \bibinfo{author}{\bibfnamefont{I.}~\bibnamefont{Cirac}},
  \bibinfo{author}{\bibfnamefont{G.~V.} \bibnamefont{Shlyapnikov}},
  \bibinfo{author}{\bibfnamefont{T.~W.} \bibnamefont{Hansch}},
  \bibnamefont{and} \bibinfo{author}{\bibfnamefont{I.}~\bibnamefont{Bloch}},
  \bibinfo{journal}{Nature} \textbf{\bibinfo{volume}{429}},
  \bibinfo{pages}{277} (\bibinfo{year}{2004}).

\bibitem[{\citenamefont{Preiss et~al.}(2015)\citenamefont{Preiss, Ma, Tai,
  Lukin, Rispoli, Zupancic, Lahini, Islam, and Greiner}}]{Preiss2015}
\bibinfo{author}{\bibfnamefont{P.~M.} \bibnamefont{Preiss}},
  \bibinfo{author}{\bibfnamefont{R.}~\bibnamefont{Ma}},
  \bibinfo{author}{\bibfnamefont{M.~E.} \bibnamefont{Tai}},
  \bibinfo{author}{\bibfnamefont{A.}~\bibnamefont{Lukin}},
  \bibinfo{author}{\bibfnamefont{M.}~\bibnamefont{Rispoli}},
  \bibinfo{author}{\bibfnamefont{P.}~\bibnamefont{Zupancic}},
  \bibinfo{author}{\bibfnamefont{Y.}~\bibnamefont{Lahini}},
  \bibinfo{author}{\bibfnamefont{R.}~\bibnamefont{Islam}}, \bibnamefont{and}
  \bibinfo{author}{\bibfnamefont{M.}~\bibnamefont{Greiner}},
  \bibinfo{journal}{Science} \textbf{\bibinfo{volume}{347}},
  \bibinfo{pages}{1229} (\bibinfo{year}{2015}), ISSN \bibinfo{issn}{0036-8075}.

\bibitem[{\citenamefont{Sachdev}(2001)}]{Sachdev2001}
\bibinfo{author}{\bibfnamefont{S.}~\bibnamefont{Sachdev}},
  \emph{\bibinfo{title}{Quantum Phase Transitions}}
  (\bibinfo{publisher}{Cambridge University Press}, \bibinfo{year}{2001}), ISBN
  \bibinfo{isbn}{9780521004541}.

\bibitem[{\citenamefont{Boll et~al.}(2016)\citenamefont{Boll, Hilker, Salomon,
  Omran, Nespolo, Pollet, Bloch, and Gross}}]{Boll2016}
\bibinfo{author}{\bibfnamefont{M.}~\bibnamefont{Boll}},
  \bibinfo{author}{\bibfnamefont{T.~A.} \bibnamefont{Hilker}},
  \bibinfo{author}{\bibfnamefont{G.}~\bibnamefont{Salomon}},
  \bibinfo{author}{\bibfnamefont{A.}~\bibnamefont{Omran}},
  \bibinfo{author}{\bibfnamefont{J.}~\bibnamefont{Nespolo}},
  \bibinfo{author}{\bibfnamefont{L.}~\bibnamefont{Pollet}},
  \bibinfo{author}{\bibfnamefont{I.}~\bibnamefont{Bloch}}, \bibnamefont{and}
  \bibinfo{author}{\bibfnamefont{C.}~\bibnamefont{Gross}},
  \bibinfo{journal}{Science} \textbf{\bibinfo{volume}{353}},
  \bibinfo{pages}{1257} (\bibinfo{year}{2016}), ISSN \bibinfo{issn}{0036-8075}.

\bibitem[{\citenamefont{Parsons et~al.}(2016)\citenamefont{Parsons, Mazurenko,
  Chiu, Ji, Greif, and Greiner}}]{Parsons2016}
\bibinfo{author}{\bibfnamefont{M.~F.} \bibnamefont{Parsons}},
  \bibinfo{author}{\bibfnamefont{A.}~\bibnamefont{Mazurenko}},
  \bibinfo{author}{\bibfnamefont{C.~S.} \bibnamefont{Chiu}},
  \bibinfo{author}{\bibfnamefont{G.}~\bibnamefont{Ji}},
  \bibinfo{author}{\bibfnamefont{D.}~\bibnamefont{Greif}}, \bibnamefont{and}
  \bibinfo{author}{\bibfnamefont{M.}~\bibnamefont{Greiner}},
  \bibinfo{journal}{Science} \textbf{\bibinfo{volume}{353}},
  \bibinfo{pages}{1253} (\bibinfo{year}{2016}), ISSN \bibinfo{issn}{0036-8075}.

\bibitem[{\citenamefont{Cheuk et~al.}(2016)\citenamefont{Cheuk, Nichols,
  Lawrence, Okan, Zhang, Khatami, Trivedi, Paiva, Rigol, and
  Zwierlein}}]{Cheuk2016}
\bibinfo{author}{\bibfnamefont{L.~W.} \bibnamefont{Cheuk}},
  \bibinfo{author}{\bibfnamefont{M.~A.} \bibnamefont{Nichols}},
  \bibinfo{author}{\bibfnamefont{K.~R.} \bibnamefont{Lawrence}},
  \bibinfo{author}{\bibfnamefont{M.}~\bibnamefont{Okan}},
  \bibinfo{author}{\bibfnamefont{H.}~\bibnamefont{Zhang}},
  \bibinfo{author}{\bibfnamefont{E.}~\bibnamefont{Khatami}},
  \bibinfo{author}{\bibfnamefont{N.}~\bibnamefont{Trivedi}},
  \bibinfo{author}{\bibfnamefont{T.}~\bibnamefont{Paiva}},
  \bibinfo{author}{\bibfnamefont{M.}~\bibnamefont{Rigol}}, \bibnamefont{and}
  \bibinfo{author}{\bibfnamefont{M.~W.} \bibnamefont{Zwierlein}},
  \bibinfo{journal}{Science} \textbf{\bibinfo{volume}{353}},
  \bibinfo{pages}{1260} (\bibinfo{year}{2016}), ISSN \bibinfo{issn}{0036-8075}.

\bibitem[{\citenamefont{Brown et~al.}(2016)\citenamefont{Brown, Mitra,
  Guardado-Sanchez, Schau{\ss}, Kondov, Khatami, Paiva, Trivedi, Huse, and
  Bakr}}]{Brown2016}
\bibinfo{author}{\bibfnamefont{P.~T.} \bibnamefont{Brown}},
  \bibinfo{author}{\bibfnamefont{D.}~\bibnamefont{Mitra}},
  \bibinfo{author}{\bibfnamefont{E.}~\bibnamefont{Guardado-Sanchez}},
  \bibinfo{author}{\bibfnamefont{P.}~\bibnamefont{Schau{\ss}}},
  \bibinfo{author}{\bibfnamefont{S.~S.} \bibnamefont{Kondov}},
  \bibinfo{author}{\bibfnamefont{E.}~\bibnamefont{Khatami}},
  \bibinfo{author}{\bibfnamefont{T.}~\bibnamefont{Paiva}},
  \bibinfo{author}{\bibfnamefont{N.}~\bibnamefont{Trivedi}},
  \bibinfo{author}{\bibfnamefont{D.~A.} \bibnamefont{Huse}}, \bibnamefont{and}
  \bibinfo{author}{\bibfnamefont{W.~S.} \bibnamefont{Bakr}}
  (\bibinfo{year}{2016}).

\bibitem[{\citenamefont{Haller et~al.}(2015)\citenamefont{Haller, Hudson,
  Kelly, Cotta, Peaudecerf, Bruce, and Kuhr}}]{Haller2015}
\bibinfo{author}{\bibfnamefont{E.}~\bibnamefont{Haller}},
  \bibinfo{author}{\bibfnamefont{J.}~\bibnamefont{Hudson}},
  \bibinfo{author}{\bibfnamefont{A.}~\bibnamefont{Kelly}},
  \bibinfo{author}{\bibfnamefont{D.~A.} \bibnamefont{Cotta}},
  \bibinfo{author}{\bibfnamefont{B.}~\bibnamefont{Peaudecerf}},
  \bibinfo{author}{\bibfnamefont{G.~D.} \bibnamefont{Bruce}}, \bibnamefont{and}
  \bibinfo{author}{\bibfnamefont{S.}~\bibnamefont{Kuhr}}, \bibinfo{journal}{Nat
  Phys} \textbf{\bibinfo{volume}{11}}, \bibinfo{pages}{738}
  (\bibinfo{year}{2015}).

\bibitem[{\citenamefont{Edge et~al.}(2015)\citenamefont{Edge, Anderson, Jervis,
  McKay, Day, Trotzky, and Thywissen}}]{Edge2015}
\bibinfo{author}{\bibfnamefont{G.~J.~A.} \bibnamefont{Edge}},
  \bibinfo{author}{\bibfnamefont{R.}~\bibnamefont{Anderson}},
  \bibinfo{author}{\bibfnamefont{D.}~\bibnamefont{Jervis}},
  \bibinfo{author}{\bibfnamefont{D.~C.} \bibnamefont{McKay}},
  \bibinfo{author}{\bibfnamefont{R.}~\bibnamefont{Day}},
  \bibinfo{author}{\bibfnamefont{S.}~\bibnamefont{Trotzky}}, \bibnamefont{and}
  \bibinfo{author}{\bibfnamefont{J.~H.} \bibnamefont{Thywissen}},
  \bibinfo{journal}{Phys. Rev. A} \textbf{\bibinfo{volume}{92}},
  \bibinfo{pages}{063406} (\bibinfo{year}{2015}).

\bibitem[{\citenamefont{Dum et~al.}(1992)\citenamefont{Dum, Parkins, Zoller,
  and Gardiner}}]{Dum1992}
\bibinfo{author}{\bibfnamefont{R.}~\bibnamefont{Dum}},
  \bibinfo{author}{\bibfnamefont{A.~S.} \bibnamefont{Parkins}},
  \bibinfo{author}{\bibfnamefont{P.}~\bibnamefont{Zoller}}, \bibnamefont{and}
  \bibinfo{author}{\bibfnamefont{C.~W.} \bibnamefont{Gardiner}},
  \bibinfo{journal}{Phys. Rev. A} \textbf{\bibinfo{volume}{46}},
  \bibinfo{pages}{4382} (\bibinfo{year}{1992}).

\bibitem[{\citenamefont{M{\o}lmer et~al.}(1993)\citenamefont{M{\o}lmer, Castin,
  and Dalibard}}]{Molmer1993}
\bibinfo{author}{\bibfnamefont{K.}~\bibnamefont{M{\o}lmer}},
  \bibinfo{author}{\bibfnamefont{Y.}~\bibnamefont{Castin}}, \bibnamefont{and}
  \bibinfo{author}{\bibfnamefont{J.}~\bibnamefont{Dalibard}},
  \bibinfo{journal}{J. Opt. Soc. Am. B} \textbf{\bibinfo{volume}{10}},
  \bibinfo{pages}{524} (\bibinfo{year}{1993}).

\bibitem[{\citenamefont{Daley}(2014)}]{Daley2014}
\bibinfo{author}{\bibfnamefont{A.~J.} \bibnamefont{Daley}},
  \bibinfo{journal}{Advances in Physics} \textbf{\bibinfo{volume}{63}},
  \bibinfo{pages}{77} (\bibinfo{year}{2014}).

\bibitem[{\citenamefont{McKay and DeMarco}(2011)}]{McKay2011}
\bibinfo{author}{\bibfnamefont{D.~C.} \bibnamefont{McKay}} \bibnamefont{and}
  \bibinfo{author}{\bibfnamefont{B.}~\bibnamefont{DeMarco}},
  \bibinfo{journal}{Reports on Progress in Physics}
  \textbf{\bibinfo{volume}{74}}, \bibinfo{pages}{054401}
  (\bibinfo{year}{2011}).

\bibitem[{\citenamefont{Weitenberg et~al.}(2011)\citenamefont{Weitenberg,
  Endres, Sherson, Cheneau, Schausz, Fukuhara, Bloch, and
  Kuhr}}]{Weitenberg2011}
\bibinfo{author}{\bibfnamefont{C.}~\bibnamefont{Weitenberg}},
  \bibinfo{author}{\bibfnamefont{M.}~\bibnamefont{Endres}},
  \bibinfo{author}{\bibfnamefont{J.~F.} \bibnamefont{Sherson}},
  \bibinfo{author}{\bibfnamefont{M.}~\bibnamefont{Cheneau}},
  \bibinfo{author}{\bibfnamefont{P.}~\bibnamefont{Schausz}},
  \bibinfo{author}{\bibfnamefont{T.}~\bibnamefont{Fukuhara}},
  \bibinfo{author}{\bibfnamefont{I.}~\bibnamefont{Bloch}}, \bibnamefont{and}
  \bibinfo{author}{\bibfnamefont{S.}~\bibnamefont{Kuhr}},
  \bibinfo{journal}{Nature} \textbf{\bibinfo{volume}{471}},
  \bibinfo{pages}{319} (\bibinfo{year}{2011}).

\bibitem[{\citenamefont{Gericke et~al.}(2008)\citenamefont{Gericke, Wurtz,
  Reitz, Langen, and Ott}}]{Gericke2008}
\bibinfo{author}{\bibfnamefont{T.}~\bibnamefont{Gericke}},
  \bibinfo{author}{\bibfnamefont{P.}~\bibnamefont{Wurtz}},
  \bibinfo{author}{\bibfnamefont{D.}~\bibnamefont{Reitz}},
  \bibinfo{author}{\bibfnamefont{T.}~\bibnamefont{Langen}}, \bibnamefont{and}
  \bibinfo{author}{\bibfnamefont{H.}~\bibnamefont{Ott}}, \bibinfo{journal}{Nat
  Phys} \textbf{\bibinfo{volume}{4}}, \bibinfo{pages}{949}
  (\bibinfo{year}{2008}).

\bibitem[{\citenamefont{Gersch and Knollman}(1963)}]{Gersch1963}
\bibinfo{author}{\bibfnamefont{H.~A.} \bibnamefont{Gersch}} \bibnamefont{and}
  \bibinfo{author}{\bibfnamefont{G.~C.} \bibnamefont{Knollman}},
  \bibinfo{journal}{Phys. Rev.} \textbf{\bibinfo{volume}{129}},
  \bibinfo{pages}{959} (\bibinfo{year}{1963}).

\bibitem[{\citenamefont{Poletti et~al.}(2013)\citenamefont{Poletti, Barmettler,
  Georges, and Kollath}}]{Poletti2013}
\bibinfo{author}{\bibfnamefont{D.}~\bibnamefont{Poletti}},
  \bibinfo{author}{\bibfnamefont{P.}~\bibnamefont{Barmettler}},
  \bibinfo{author}{\bibfnamefont{A.}~\bibnamefont{Georges}}, \bibnamefont{and}
  \bibinfo{author}{\bibfnamefont{C.}~\bibnamefont{Kollath}},
  \bibinfo{journal}{Phys. Rev. Lett.} \textbf{\bibinfo{volume}{111}},
  \bibinfo{pages}{195301} (\bibinfo{year}{2013}).

\bibitem[{\citenamefont{White}(1992)}]{White1992}
\bibinfo{author}{\bibfnamefont{S.~R.} \bibnamefont{White}},
  \bibinfo{journal}{Phys. Rev. Lett.} \textbf{\bibinfo{volume}{69}},
  \bibinfo{pages}{2863} (\bibinfo{year}{1992}).

\bibitem[{\citenamefont{Verstraete et~al.}(2008)\citenamefont{Verstraete, Murg,
  and Cirac}}]{Verstraete2008}
\bibinfo{author}{\bibfnamefont{F.}~\bibnamefont{Verstraete}},
  \bibinfo{author}{\bibfnamefont{V.}~\bibnamefont{Murg}}, \bibnamefont{and}
  \bibinfo{author}{\bibfnamefont{J.~I.} \bibnamefont{Cirac}},
  \bibinfo{journal}{Adv. Phys.} \textbf{\bibinfo{volume}{57}},
  \bibinfo{pages}{143 } (\bibinfo{year}{2008}).

\bibitem[{\citenamefont{Schollw{\"o}ck}(2011)}]{Schollwock2011}
\bibinfo{author}{\bibfnamefont{U.}~\bibnamefont{Schollw{\"o}ck}},
  \bibinfo{journal}{Annals of Physics} \textbf{\bibinfo{volume}{326}},
  \bibinfo{pages}{96 } (\bibinfo{year}{2011}), ISSN \bibinfo{issn}{0003-4916},
  \bibinfo{note}{january 2011 Special Issue}.

\bibitem[{\citenamefont{Vidal}(2003)}]{Vidal2003}
\bibinfo{author}{\bibfnamefont{G.}~\bibnamefont{Vidal}},
  \bibinfo{journal}{Phys. Rev. Lett.} \textbf{\bibinfo{volume}{91}},
  \bibinfo{pages}{147902} (\bibinfo{year}{2003}).

\bibitem[{\citenamefont{Pirvu et~al.}(2010)\citenamefont{Pirvu, Murg, Cirac,
  and Verstraete}}]{Pirvu2010}
\bibinfo{author}{\bibfnamefont{B.}~\bibnamefont{Pirvu}},
  \bibinfo{author}{\bibfnamefont{V.}~\bibnamefont{Murg}},
  \bibinfo{author}{\bibfnamefont{J.~I.} \bibnamefont{Cirac}}, \bibnamefont{and}
  \bibinfo{author}{\bibfnamefont{F.}~\bibnamefont{Verstraete}},
  \bibinfo{journal}{New Journal of Physics} \textbf{\bibinfo{volume}{12}},
  \bibinfo{pages}{025012} (\bibinfo{year}{2010}).

\bibitem[{\citenamefont{Verstraete et~al.}(2004)\citenamefont{Verstraete,
  Garc\'{\i}a-Ripoll, and Cirac}}]{Verstraete2004}
\bibinfo{author}{\bibfnamefont{F.}~\bibnamefont{Verstraete}},
  \bibinfo{author}{\bibfnamefont{J.~J.} \bibnamefont{Garc\'{\i}a-Ripoll}},
  \bibnamefont{and} \bibinfo{author}{\bibfnamefont{J.~I.} \bibnamefont{Cirac}},
  \bibinfo{journal}{Phys. Rev. Lett.} \textbf{\bibinfo{volume}{93}},
  \bibinfo{pages}{207204} (\bibinfo{year}{2004}).

\bibitem[{\citenamefont{Schollw\"ock}(2005)}]{Schollwock2005}
\bibinfo{author}{\bibfnamefont{U.}~\bibnamefont{Schollw\"ock}},
  \bibinfo{journal}{Rev. Mod. Phys.} \textbf{\bibinfo{volume}{77}},
  \bibinfo{pages}{259} (\bibinfo{year}{2005}).

\bibitem[{\citenamefont{Daley et~al.}(2004)\citenamefont{Daley, Kollath,
  Schollw\"ock, and Vidal}}]{Daley2004}
\bibinfo{author}{\bibfnamefont{A.~J.} \bibnamefont{Daley}},
  \bibinfo{author}{\bibfnamefont{C.}~\bibnamefont{Kollath}},
  \bibinfo{author}{\bibfnamefont{U.}~\bibnamefont{Schollw\"ock}},
  \bibnamefont{and} \bibinfo{author}{\bibfnamefont{G.}~\bibnamefont{Vidal}},
  \bibinfo{journal}{J. Stat. Mech.} p. \bibinfo{pages}{P04005}
  (\bibinfo{year}{2004}).

\bibitem[{\citenamefont{Lieb and Robinson}(1972)}]{Lieb1972}
\bibinfo{author}{\bibfnamefont{E.}~\bibnamefont{Lieb}} \bibnamefont{and}
  \bibinfo{author}{\bibfnamefont{D.}~\bibnamefont{Robinson}},
  \bibinfo{journal}{Commun. Math. Phys.} \textbf{\bibinfo{volume}{28}},
  \bibinfo{pages}{251} (\bibinfo{year}{1972}).

\bibitem[{\citenamefont{Islam et~al.}(2015)\citenamefont{Islam, Ma, Preiss,
  Eric~Tai, Lukin, Rispoli, and Greiner}}]{Islam2015}
\bibinfo{author}{\bibfnamefont{R.}~\bibnamefont{Islam}},
  \bibinfo{author}{\bibfnamefont{R.}~\bibnamefont{Ma}},
  \bibinfo{author}{\bibfnamefont{P.~M.} \bibnamefont{Preiss}},
  \bibinfo{author}{\bibfnamefont{M.}~\bibnamefont{Eric~Tai}},
  \bibinfo{author}{\bibfnamefont{A.}~\bibnamefont{Lukin}},
  \bibinfo{author}{\bibfnamefont{M.}~\bibnamefont{Rispoli}}, \bibnamefont{and}
  \bibinfo{author}{\bibfnamefont{M.}~\bibnamefont{Greiner}},
  \bibinfo{journal}{Nature} \textbf{\bibinfo{volume}{528}}, \bibinfo{pages}{77}
  (\bibinfo{year}{2015}).

\bibitem[{\citenamefont{Daley et~al.}(2012)\citenamefont{Daley, Pichler,
  Schachenmayer, and Zoller}}]{Daley2012}
\bibinfo{author}{\bibfnamefont{A.~J.} \bibnamefont{Daley}},
  \bibinfo{author}{\bibfnamefont{H.}~\bibnamefont{Pichler}},
  \bibinfo{author}{\bibfnamefont{J.}~\bibnamefont{Schachenmayer}},
  \bibnamefont{and} \bibinfo{author}{\bibfnamefont{P.}~\bibnamefont{Zoller}},
  \bibinfo{journal}{Phys. Rev. Lett.} \textbf{\bibinfo{volume}{109}},
  \bibinfo{pages}{020505} (\bibinfo{year}{2012}).

\bibitem[{\citenamefont{Pichler et~al.}(2013)\citenamefont{Pichler, Bonnes,
  Daley, L{\"a}uchli, and Zoller}}]{Pichler2013b}
\bibinfo{author}{\bibfnamefont{H.}~\bibnamefont{Pichler}},
  \bibinfo{author}{\bibfnamefont{L.}~\bibnamefont{Bonnes}},
  \bibinfo{author}{\bibfnamefont{A.~J.} \bibnamefont{Daley}},
  \bibinfo{author}{\bibfnamefont{A.~M.} \bibnamefont{L{\"a}uchli}},
  \bibnamefont{and} \bibinfo{author}{\bibfnamefont{P.}~\bibnamefont{Zoller}},
  \bibinfo{journal}{New Journal of Physics} \textbf{\bibinfo{volume}{15}},
  \bibinfo{pages}{063003} (\bibinfo{year}{2013}).

\bibitem[{\citenamefont{Schreiber et~al.}(2015)\citenamefont{Schreiber,
  Hodgman, Bordia, L{\"u}schen, Fischer, Vosk, Altman, Schneider, and
  Bloch}}]{Schreiber2015}
\bibinfo{author}{\bibfnamefont{M.}~\bibnamefont{Schreiber}},
  \bibinfo{author}{\bibfnamefont{S.~S.} \bibnamefont{Hodgman}},
  \bibinfo{author}{\bibfnamefont{P.}~\bibnamefont{Bordia}},
  \bibinfo{author}{\bibfnamefont{H.~P.} \bibnamefont{L{\"u}schen}},
  \bibinfo{author}{\bibfnamefont{M.~H.} \bibnamefont{Fischer}},
  \bibinfo{author}{\bibfnamefont{R.}~\bibnamefont{Vosk}},
  \bibinfo{author}{\bibfnamefont{E.}~\bibnamefont{Altman}},
  \bibinfo{author}{\bibfnamefont{U.}~\bibnamefont{Schneider}},
  \bibnamefont{and} \bibinfo{author}{\bibfnamefont{I.}~\bibnamefont{Bloch}},
  \bibinfo{journal}{Science} \textbf{\bibinfo{volume}{349}},
  \bibinfo{pages}{842} (\bibinfo{year}{2015}), ISSN \bibinfo{issn}{0036-8075}.

\end{thebibliography}

\end{document}